\documentclass[aps,prl,twocolumn,reprint,showpacs,superscriptaddress,footinbib,groupedaddress]{revtex4-1}
\usepackage{dsfont}
\usepackage{amsmath}
\usepackage{graphicx}
\usepackage[utf8]{inputenc}
\usepackage{ulem} %added to be able to cross text out
\usepackage{units} % for using nicefrac in texts
\usepackage{version} % for comment
\usepackage{color}
\usepackage{colortbl}
\usepackage{xcolor}  % for writting parts in color, e.g. in red
\usepackage{fancyref}
\usepackage[colorlinks=true,citecolor=black,urlcolor=blue,linkcolor=black]{hyperref} % choice of color for hyperref links

\usepackage{multirow}
\usepackage{array}
\usepackage{booktabs}
\usepackage{longtable}
\newcommand{\ms}{\ \text{ms}}

\newcommand{\Hz}{\ \text{Hz}}

\newcommand{\sqrtHz}{/\sqrt{\text{Hz}}}

\newcommand{\nrads}{\ \text{nrad.s}^{-1}}

\newcommand{\beq}{\begin{equation}}
\newcommand{\eeq}{\end{equation}}

\begin{document}
\graphicspath{{Figures/}}
\preprint{AIP/123-QED}

%\title{Continuous Cold Atom Gyroscope with $1\nrads$ stability}
\title{Continuous Cold-atom Inertial Sensor with $1\nrads$ Rotation Stability}

\author{I. Dutta}
\author{D. Savoie}
\author{B. Fang}
\author{B. Venon}
\author{C.L. Garrido Alzar}
\author{R. Geiger}
\email{remi.geiger@obspm.fr}
\author{A. Landragin}
\email{arnaud.landragin@obspm.fr}

\affiliation{LNE-SYRTE, Observatoire de Paris, PSL Research University, CNRS, Sorbonne Universit\'es, UPMC Univ. Paris 06,  61 avenue de l'Observatoire, 75014 Paris, France}

\date{\today}

\begin{abstract}
We report the operation of a cold-atom inertial sensor which continuously captures the rotation signal.
Using a joint interrogation scheme, where we simultaneously prepare a cold-atom source and operate an atom interferometer (AI) enables us to eliminate the dead times.
We show that such continuous operation improves the short-term sensitivity of AIs, and demonstrate a  rotation sensitivity of $100\nrads\sqrtHz$ in a cold-atom gyroscope of  $11 \ \text{cm}^2$ Sagnac area. 
We also demonstrate  a rotation stability of $1 \nrads$ at $10^4 \ \text{s}$ of integration time, which establishes the record for atomic gyroscopes.
The continuous operation of cold-atom inertial sensors will enable to benefit from the full sensitivity potential of large area AIs, determined by the quantum noise limit.

\end{abstract}

\pacs{03.75.Dg, 37.25.+k}
% for PACS : http://www.aip.org/publishing/pacs/pacs-alphabetical-index
% 06.30.Ft, 95.55.Sh = clocks
% 37.25.+k, 03.75.Dg = atom interferometry

%\keywords{}

\maketitle

\label{introduction}

Over the past two decades, important progress in cold-atom physics has established atom interferometry  as a unique tool for precision measurements of time and frequency and of gravito-inertial effects. Atom interferometry addresses various applications ranging from precision measurements of fundamental constants \cite{Bouchendira2011,Rosi2014}, to inertial navigation \cite{Canuel2006,Geiger2011,stockton_absolute_2011}, to geophysics and geodesy \cite{Gillot2014,Geiger2015,Freier2015,RosiMeasurement2015} and has been proposed for gravitational wave detection \cite{Dimopoulos2008PRD,Chaibi2016}. 
New techniques are being developed to improve the potential of atom interferometers (AIs), such as large momentum transfer beam splitters \cite{Clade2009,Chiow2011}, long interrogation times in tall vacuum chambers \cite{Dickerson2013}, microgravity platforms \cite{Geiger2011,Muntinga2013}, or operation of AIs with  ultracold atomic sources \cite{Debs2011}. Advanced detection and atom preparation methods have moreover been proposed and demonstrated  to go beyond the quantum projection noise in AIs \cite{Leroux2010,Hosten2016}.
However, benefiting from these new techniques to fully exploit the potential of AIs requires to handle the problem of dead times between successive measurements occurring in  cold-atom sensors.

Dead times in AIs originate from the preparation of the atomic source prior to the entrance in the interferometric zone and to the detection of the atoms at the AI output. The inertial information during these preparation and detection periods is lost.
Dead times, for example, strongly mitigate the possibility to realize inertial measurement units (IMUs) based on AIs \cite{jekeli_navigation_2005}.
In addition, the sequential operation of AIs leads to inertial noise aliasing, which  degrades the AI sensitivity in the presence of dead times. 
This reduces the performance of AIs of potentially  high sensitivities \cite{Dickerson2013}.
High data rate interferometers using  recapture methods have been reported to partially overcome the problem of dead times but at the cost of strong reduction of sensitivity \cite{mcguinness_high_2012}.
The inertial noise aliasing in  AIs can be alleviated by using auxiliary sensors of large bandwidth \cite{lautier_hybridizing_2014}, but this limits the sensitivity  during the dead time period to that of the auxiliary sensor.
Continuous operation (i.e. without dead times)  is therefore a key point to benefit from the full potential of atom interferometry.

In this letter, we report the first continuous operation of a cold-atom inertial sensor. We demonstrate such  operation in an AI gyroscope which features a Sagnac area of $11 \ \text{cm}^2$, representing a $27$-fold increase with respect to  previous experiments \cite{berg_composite-light-pulse_2015}.
The continuous operation improves the short-term sensitivity of the gyroscope, which we illustrate by demonstrating a  rotation sensitivity of $100\nrads\sqrtHz$. Moreover, we show that the continuous operation does not affect the long-term sensitivity potential of AIs and report a  stability of $1\nrads$ after $10^4 \ \text{s}$ of integration time.

\begin{figure}[!h]
\includegraphics[width=\linewidth]{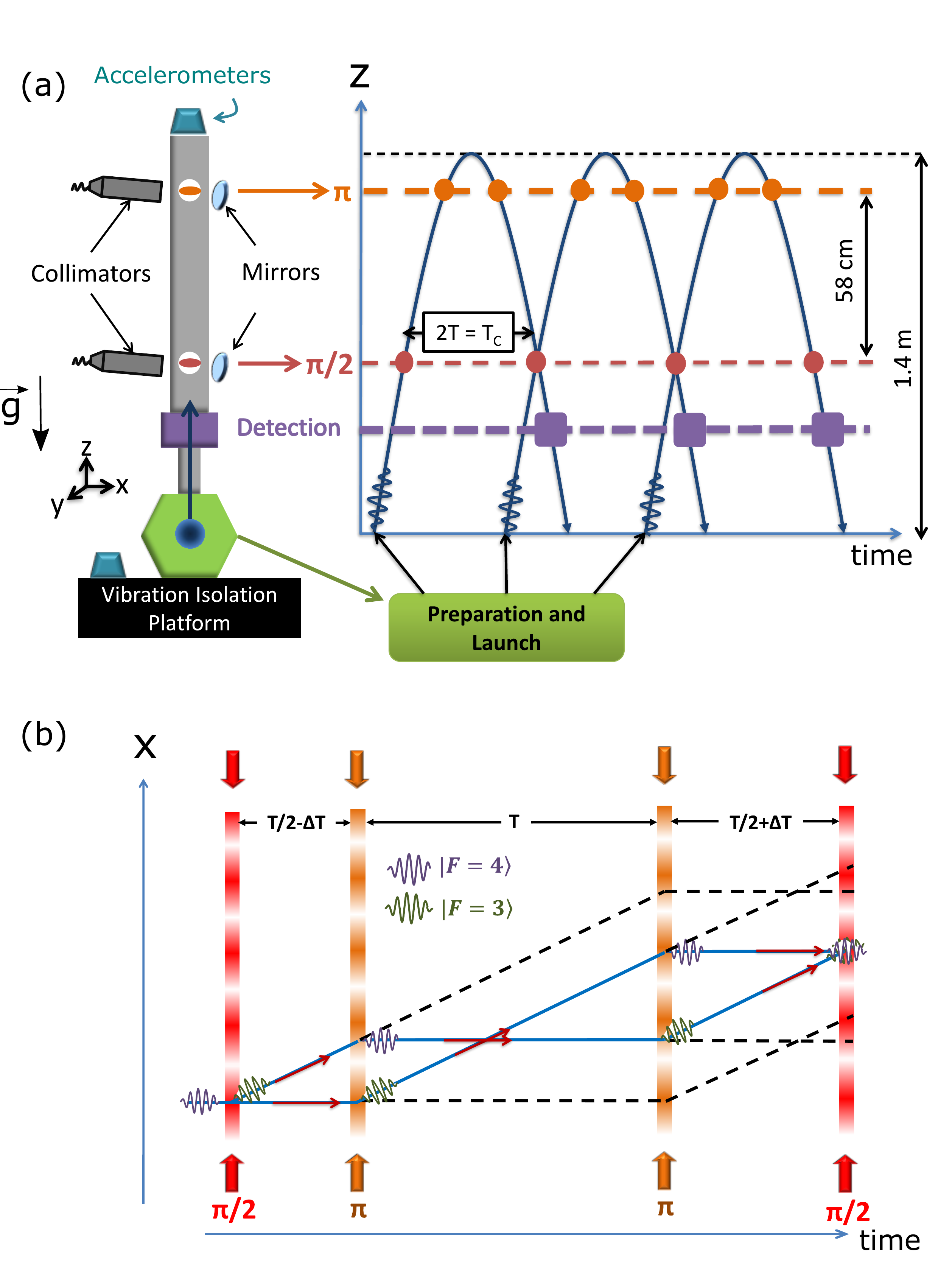}
\caption{\label{fig:contsch} 
(Color online)
 (a) Schematic and operation principle of the continuous cold-atom gyroscope. Continuous measurement is performed with a joint interrogation sequence where the bottom $\pi/2$ pulse is shared between the clouds entering and exiting the interrogation region. 
(b) Space-time  diagram of the  four-pulse AI. We introduce a time asymmetry of $\Delta T$ to avoid the recombination of parasitic interferometers resulting from the imperfect $\pi$ pulses.
The gyroscope measures rotation rate along the $y$ direction, i.e. perpendicular to the AI area.
}
\end{figure}

\label{principle}
% In this paragraph we explain the general principle of the gyroscope and of the joint measurement sequence.
The principle of the experiment is sketched in Fig.~\ref{fig:contsch}. We realize a light-pulse AI using two counter-propagating Raman beams which couple the $|F=3,m_F=0\rangle$ and $|F=4,m_F=0\rangle$ clock states of Cesium atoms.
According to the Sagnac effect \cite{Sagnac1913,barrett_sagnac_2014}, the rotation sensitivity of the AI is proportional to the area enclosed by the 2 arms. Our AI gyroscope is based on a fountain configuration with four Raman pulses to create a folded geometry thanks to gravity \cite{Canuel2006}.
Similar folded geometries can be obtained in trapped atom interferometers \cite{Wu2007}.
The four pulse fountain configuration allows us to increase the interferometric area up to $11 \ \text{cm}^2$ and  leads to zero DC sensitivity to acceleration. The rotation induced phase shift $\Phi_{\Omega}$ is given by
\begin{equation}\label{eqn:rotphase4p}
\Phi_{\Omega}= \frac{1}{2}\vec{k}_{\mathrm{eff}}\cdot \left(\vec{g}\times\vec{\Omega}\right) T^3,
\end{equation}
where $\vec{k}_{\mathrm{eff}}$ is the two-photon momentum transfer, $\vec{g}$ is the acceleration due to gravity, $\vec{\Omega}$ is the rotation rate and $T$ is half the interferometric time. 
Following atom juggling methods initially introduced to measure collisional shifts in fountain clocks \cite{Legere1998}, we implement a sequence of joint interrogation of successive atom clouds as described in \cite{Meunier2014}, see Fig.\ref{fig:contsch}(a). Experimentally, the joint operation is obtained by using the same $\pi/2$ Raman pulse for the clouds entering  and exiting the AI zone. Thus, the experiment cycle time $T_c$ equals the AI interrogation time $2T$.
%The joint operation implies to trap a cloud of atoms in the bottom part of the chamber, while another atom cloud is in the AI or detection region.

\label{experiment}
% In this paragraph we give the main details of the experiment.
Cesium atoms loaded from a 2D Magneto-Optical Trap (MOT) are trapped and cooled in a 3D-MOT during $200 \ms$. We launch $2\times 10^7$ atoms   vertically at  a speed of $5.0 \ \text{m.s}^{-1}$ using moving molasses  with a (3D) cloud  temperature of  $1.2 \ \mu\text{K}$.  
Light pulse interferometry is realized using two phase-locked Raman lasers  which couple the Cesium clock  states characterized by an hyperfine splitting corresponding to  $9.192 \ \text{GHz}$.
The Raman lasers are sent to the atoms through two optical windows separated by 58 cm, yielding an interrogation time  $2T=800 \ms$. 
 We use Raman beams with $1/e^2$ diameter equal to $40 \ \text{mm}$ and  100 mW of total power.
After the MOT and prior to the interrogation, $2\times 10^6$ atoms are prepared in the $|F=4,m_F=0\rangle$ state. The AI output signal is determined by the probability of transition from the $F=4$ to the $F=3$ state, which is experimentally realized using fluorescence detection of the two levels after the AI light-pulse sequence.

\label{alignement}
We lift the degeneracy between the two $\pm \hbar k_{\mathrm{eff}}$ transitions \cite{leveque_enhancing_2009} by tilting the  Raman beams  by an angle of inclination $\theta=3.81^\mathrm{o}$ (Fig.~\ref{fig:contsch}(a)). 
Large area AIs require precise parallelism of the interrogation beams in order for the two  paths to recombine within the coherence length of the cold atoms at the interferometer output \cite{kellogg_longitudinal_2007}. We implement a generic protocol to meet the required beam alignment  of the Raman beams in the vertical ($z$) and  horizontal ($y$) directions. 
For the $z$ direction, we first measure the two beam angles using Doppler spectroscopy, which determines the parallelism with a precision of  $20 \ \mu$rad.
We then operate two 3-pulse  AI accelerometers at the bottom and  top Raman beam positions with an interrogation time of 60 ms to measure the projection of gravity on the beam directions, which allows us to reach a precision of $5\ \mu$rad.
To adjust the horizontal ($y$) parallelism, we optimize the contrast of a Ramsey-Bord\'e AI using the bottom and top Raman beams as described in Ref.~\cite{tackmann_self-alignment_2012},  and reach a parallelism precision of $200 \ \mu\text{rad}$. 
With this  protocol, we achieve a contrast of  $4 \ \%$ in the continuous AI at  $2T= 800 \ms$, mainly limited by inhomogeneities of the Rabi frequency over the atom cloud extension. 
For this value of contrast, the AI phase noise due to detection noise amounts to $400 \ \text{mrad}\sqrtHz$ and was estimated with the method described in \cite{Geiger2011}. 
The detection noise level is  limited by stray light in the fluorescence detection system and was  measured independently without atoms in the interferometer.
The limitations associated with joint operation (mainly light shifts and contrast reduction due to scattered light by the MOT) have been described in \cite{Meunier2014}, together with mitigation strategies.

\label{vibration noise rejection}
% In this paragraph we explain how we reject the vibration noise
% start by saying that the AI has such a large scale factor that the effect of vibration noise must be strongly mitigated
The AI output signal $P$ is  determined  by the Earth rotation rate, the vibration noise and the non-inertial noise. We write it as
\begin{equation}\label{eqn:probnoise}
P = P_0 + A\cos \left(\Phi_{\Omega}+\delta\Phi_{\mathrm{vib}} + \delta\Phi_0\right),
\end{equation}
where $P_0$ is the offset of the interferometric signal, $A$ is the fringe amplitude, $\Phi_{\Omega}$ is the rotation phase, $\delta\Phi_{\mathrm{vib}}$  the vibration phase noise and $\delta\Phi_0$ the non-inertial phase noise (e.g. Raman laser phase, light shift). Increasing the AI area necessarily comes at the expense of more sensitivity to the vibration noise, $\delta\Phi_{\mathrm{vib}}$, which has to be reduced to extract the rotation signal, $\Phi_{\Omega}$. 
The experiment is mounted on a vibration isolation platform  to reduce the effect of vibration noise  above $\sim 1\Hz$ to an rms AI phase noise of about $2.5\ \text{rad}$. As the vibration noise spans more than one interferometric fringe, information from additional inertial sensors is necessary to recover the signal.

\begin{figure}[!bth]
\includegraphics[width=\linewidth]{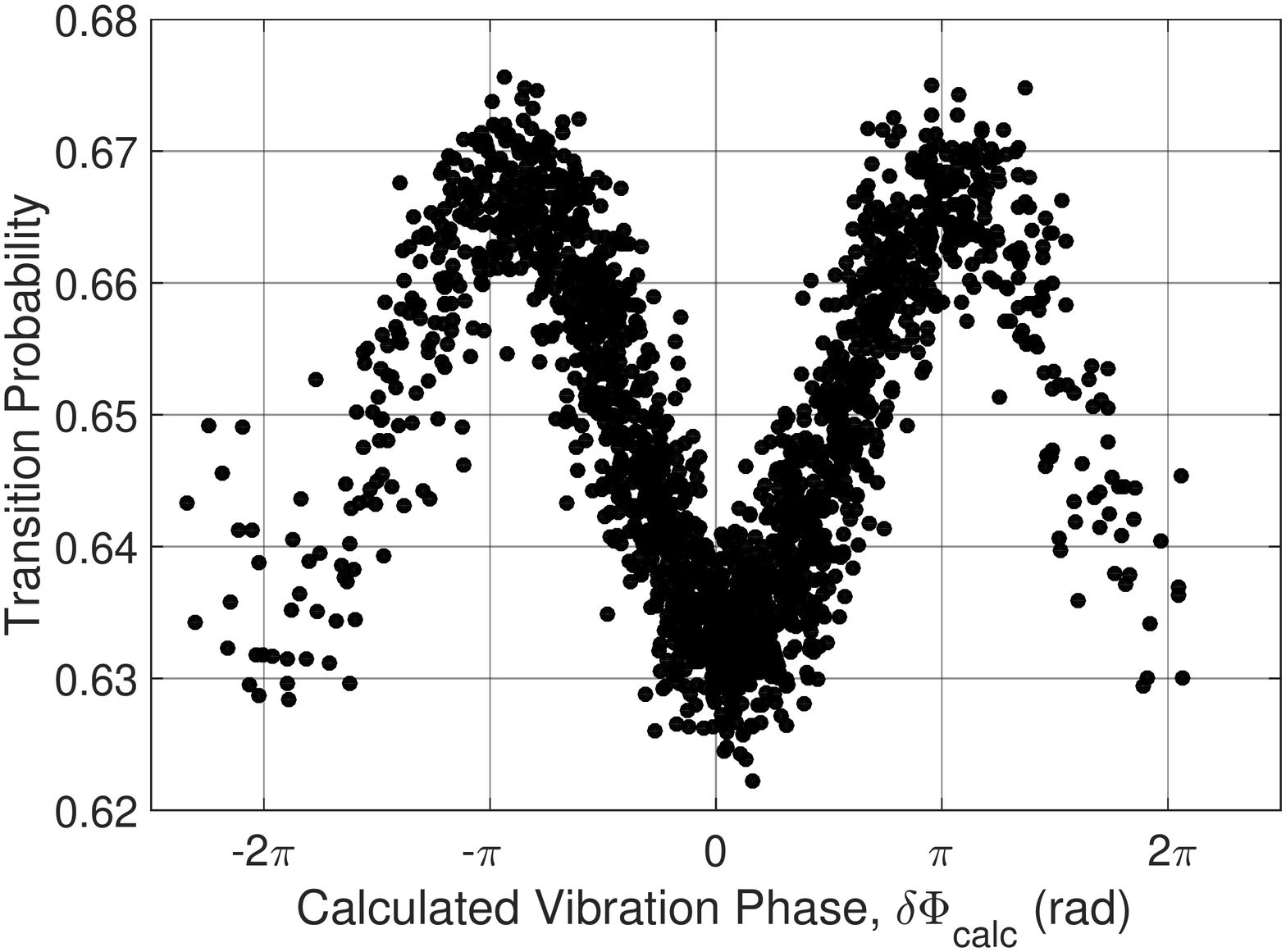}
\caption{\label{fig:contcorr}
Correlation between the AI signal  and the vibration phase calculated from the signal of auxiliary accelerometers. The AI interrogation time is $2T = 800 \ms$.
}
\end{figure}

We further reduce the vibration noise by means of auxiliary sensors which record the acceleration noise of the setup \cite{merlet_operating_2009}. We mount two commercial accelerometers (model Titan from Nanometrics) on the top and bottom of the experimental structure (see Fig.~\ref{fig:contsch}(a)), and compute the expected vibration phase $\delta\Phi_{\mathrm{calc}}$ using the four-pulse AI transfer function. 
Fig.~\ref{fig:contcorr} shows the correlation between the AI output signal, $P$, and the  phase  $\delta\Phi_{\mathrm{calc}}$ calculated from the weighted average of the two accelerometers. 
As the correlation function is non-linear, we use the method described in \cite{merlet_operating_2009} to extract the rotation rate sensitivity of the interferometer. We divide the total data set in  packets of 20 data points and fit a sinusoid to extract the offset phase and hence the rotation rate $\Omega$. This procedure yields a short-term sensitivity of $450 \ \text{mrad}\sqrtHz$, equivalent to rejecting the vibration noise by  a factor 5. 
The  rejection efficiency is limited by the  detection noise level  which currently bounds the short-term sensitivity of the AI.

\begin{figure}[!h]
\includegraphics[width=\linewidth]{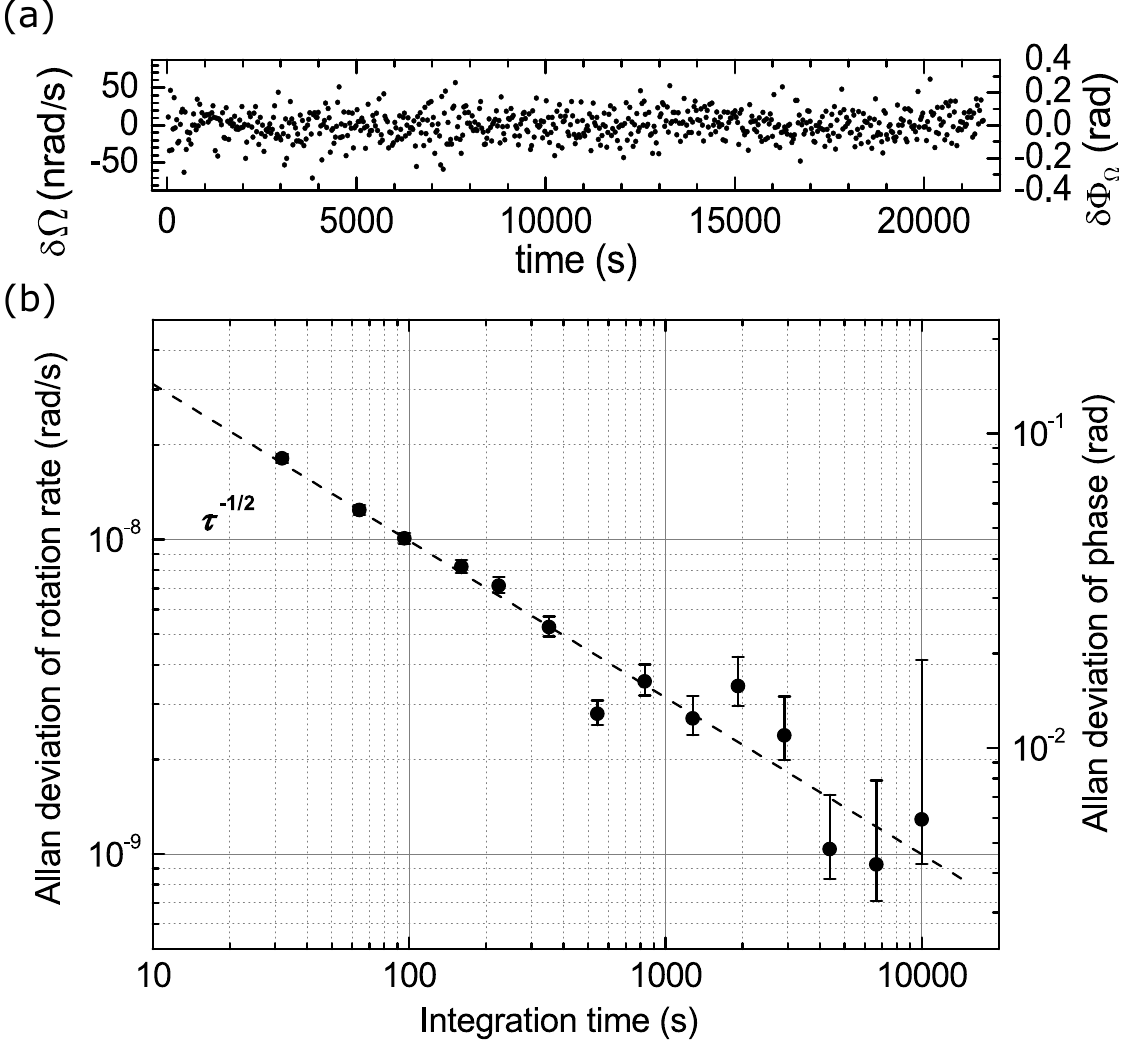}
\caption{\label{fig:contallan}
(a) Temporal variation of the rotation rate around its mean value. Each point is obtained from the combination of the  two phase measurements extracted from correlation fringes (as shown in Fig.~\ref{fig:contcorr}) involving 20 data points for each of the two opposite Raman wavevectors $+k_{\text{eff}}$ and $-k_{\text{eff}}$. (b) Allan deviation of the gyroscope sensitivity. The dashed line is a guide to the eye illustrating the $\tau^{-1/2}$ scaling.  The error bars represent the $68\%$ confidence intervals.
}
\end{figure}

Figure \ref{fig:contallan}(a) shows an uninterrupted operation of the continuous cold-atom gyroscope over more than $20000 \ \text{s}$.
The Allan deviation of the rotation rate sensitivity  is shown in Fig.~\ref{fig:contallan}(b). 
We achieve a short-term sensitivity of $100\nrads\sqrtHz$, which establishes the best performance to date for cold-atom gyroscopes \cite{berg_composite-light-pulse_2015}, and  represents an improvement of more than 30 compared to previous 4-pulse gyroscopes  \cite{Canuel2006,stockton_absolute_2011}.
We compared the operation of the gyroscope in normal and continuous modes and observed a sensitivity improvement of $\simeq 1.4$.  This is consistent with the expected value of $[T_c^{(n)}/2T]^{-1/2}$ where $T_c^{(n)}=2T+T_D$ is the cycle time in normal mode with a dead time $T_D\simeq 0.8 \ \text{s}$.

The stability of the rotation rate measurement improves as $\tau^{-1/2}$ and reaches $1 \nrads$ at $10000 \ \text{s}$ of integration time.
This represents the state of the art for atomic gyroscopes \cite{Durfee2006} (see \cite{barrett_sagnac_2014} for a recent review) and a more than 10-fold  improvement compared to previous cold-atom gyroscopes \cite{gauguet_characterization_2009, berg_composite-light-pulse_2015}. 
The long-term stability of our gyroscope is a direct consequence of the large  Sagnac area: the AI scale factor in our folded four-pulse geometry scales as $T^3$ when the  instabilities linked to fluctuations of the atom cloud trajectories  and identified as limits in previous experiments \cite{gauguet_characterization_2009,berg_composite-light-pulse_2015} scale as $T$. Their impact is thus reduced in our long-$T$ interferometer. 
We further eliminate the effect of drifts in one-photon light shift originating from drifts of the power ratio of the Raman lasers. This is accomplished by alternating measurements with $\pm k_{\mathrm{eff}}$ momentum transfer and combining the  fitted phase values obtained from the 20-points correlation data sets.

To avoid the interference of parasitic interferometers originating from the imperfect $\pi/2$ and $\pi$ pulses, we introduce a time asymmetry  of $\Delta T=300\ \mu\text{s}$  in the Raman pulse sequence \cite{stockton_absolute_2011}, see Fig~\ref{fig:contsch}(b). The asymmetry  introduces a sensitivity to DC acceleration given by $\Phi_{DC} = 2k_{\mathrm{eff}} T\Delta T g\sin\theta $. Fluctuations of the angle of inclination of the Raman beams by $\delta \theta$  would result in fluctuations of the AI phase $\Phi_{DC}$. To minimize these fluctuations, we stabilize the vibration isolation platform by measuring the tilt of the experiment  and using its signal to compensate the tilt variation via a current-controlled magnetic actuator. 
We stabilize $\delta\theta$ at the level of $3\times 10^{-8} \ \text{rad}$, ensuring long-term stabilization of $\Phi_{DC}$ below $0.3 \nrads$ after $2000 \ \text{s}$ of integration. % Drifts due to tidal variations of $g$ are negligible in the present study. 
Moreover, we alternated measurements with $\pm \Delta T$ and did not observe any effect on the rotation signal, as expected.
The tilt in the $y$ direction was measured to drift by less than $10 \ \mu\text{rad}$, yielding a negligible phase drift due to a different projection of the rotation vector on the interferometer area.

%\paragraph{Discussion.}
\label{discussion}

Our results represent record inertial sensitivities in a Sagnac AI. We emphasize that such performances were obtained, for the first time, without loss of information on the inertial signal thanks to the joint operation of the interferometer.
In our setup, the sensitivity is currently limited by the detection noise, yielding a $\tau^{-1/2}$ scaling of the rotation stability. 
Improving the contrast of our AI (e.g. with more powerful and larger Raman beams) and reducing the stray light in our current detection system would result in a lower detection noise limit. In that case, the continuous operation would offer the possibility to efficiently average the vibration noise as $\tau^{-1}$ as a result of noise correlations between successive measurements. 
Such  scaling of the sensitivity has been  demonstrated in clock configurations to  average the local oscillator noise  \cite{biedermann_zero-dead-time_2013,Meunier2014}. 
The continuous operation which we demonstrated here   will then enable to quickly reach the quantum  projection noise (or Heisenberg) limit in large area AIs.
Assuming a vibration noise averaging as $\tau^{-1}$, a quantum projection noise limited detection with $10^6$ atoms and a $20\%$ interferometer contrast, a rotation sensitivity below $1\times 10^{-10} \ \text{rad.s}^{-1}$ in few $100 \ \text{s}$ is thus accessible  with our setup. 

If we assume negligible detection noise, observing the $\tau^{-1}$ scaling would require to operate the AI in its linear region, i.e.  around mid-fringe.
Otherwise, the loss of inertial sensitivity, which occurs when approaching the top and bottom of the fringe, prevents from observing the $\tau^{-1}$ scaling (see Fig.~S1 in the Supplemental Material for a simulation).
Mid-fringe operation can, for example, be achieved by a real time compensation of vibrations with a feedback to the Raman laser phase \cite{lautier_hybridizing_2014}.

The sensitivity reached by our instrument allows us to foresee applications in geodesy and geophysics. 
High rotation rate sensitivity combined with the large bandwidth obtained by continuous operation and the multiple-joint technique \cite{Meunier2014} would allow, for instance, the detection of the rotational signatures of seismic signals that cover a wide range of rotation rates from $10^{-14}$ rad/s to 1 rad/s with typical signal frequencies in the range of few mHz to tens of Hz \cite{Schreiber2013}. Moreover, signals due to Earth tides, polar motion and ocean loading could be accessible with our device.

The continuous operation which we demonstrated here paves the way to inertial navigation based on AIs, by fully exploiting the sensitivity and long-term stability of atomic sensors  without loss of information \cite{jekeli_navigation_2005}.
Finally, the continuous operation will benefit to fundamental physics experiments     with AIs, in particular when looking for time varying signals such as in gravitational wave detection \cite{Dimopoulos2008PRD,Chaibi2016}.

We acknowledge the financial support from D\'el\'egation G\'en\'erale de l'Armement (DGA contract No. 2010.34.0005), Centre National d'Etudes Saptiales (CNES), Institut Francilien de Recherche sur les Atomes Froids (IFRAF), the Action Sp\'ecifique du CNRS Gravitation, R\'ef\'erences, Astronomie et M\'etrologie (GRAM) and Ville de Paris (project HSENS-MWGRAV). I.D. was supported by CNES and FIRST-TF (ANR-10-LABX-48-01), D.S. by DGA and B.F. by FIRST-TF. We thank M. Meunier, T. L\'ev\`eque and D. Holleville for  contributions to the experimental setup, and F. Pereira Dos Santos for careful reading of the manuscript. 

\bibliographystyle{apsrev4-1}
\bibliography{ContiGyro}

% *****************************************
% *****************************************
% *****************************************
\end{document}